\documentclass[twocolumn,pre,amsmath,amssymb]{revtex4}

\pdfoutput=1
\usepackage{graphicx}      
\usepackage{bm}            
\usepackage{hyperref}      
\hypersetup{colorlinks,citecolor=black,filecolor=black,linkcolor=black,urlcolor=black}

\begin{document}

\title{Phase oscillators with global sinusoidal coupling evolve by M\"{o}bius group action}

\author{Seth A. Marvel\footnotemark[2]}
\email{sam255@cornell.edu}

\author{Renato E. Mirollo\footnotemark[3]}

\author{Steven H. Strogatz\footnotemark[2]}

\affiliation{\vspace{6 pt} \footnotemark[2]Center for Applied Mathematics, Cornell University, Ithaca, New York 14853 \vspace{1 pt} \\
\footnotemark[3]Department of Mathematics, Boston College, Chestnut Hill, Massachusetts 02167}

\begin{abstract}

Systems of $N$ identical phase oscillators with global sinusoidal coupling are known to display low-dimensional dynamics.  Although this phenomenon was first observed about 20 years ago, its underlying cause has remained a puzzle.  Here we expose the structure working behind the scenes of these systems, by proving that the governing equations  are generated by the action of the M\"{o}bius group, a three-parameter subgroup of fractional linear transformations that map the unit disc to itself.   When there are no auxiliary state variables, the group action partitions the $N$-dimensional state space  into three-dimensional invariant manifolds (the group orbits).  The $N - 3$ constants of motion associated with this foliation are the $N - 3$ functionally independent cross ratios of the oscillator phases.  No further reduction is possible, in general; numerical experiments on models of Josephson junction arrays suggest that the invariant manifolds often contain three-dimensional regions of neutrally stable chaos.

\end{abstract}

\maketitle

\textbf{Large arrays of coupled limit-cycle oscillators have been used to model diverse systems in physics, biology, chemistry, engineering and social science.  The special case of phase oscillators coupled all-to-all through sinusoidal interactions has attracted mathematical interest because of its analytical tractability.  About 20 years ago, numerical experiments revealed that these systems display an exceptionally simple form of collective behavior: for all $N \geq 3$, where $N$ is the number of oscillators, all trajectories are confined to manifolds with $N-3$ fewer dimensions than the state space itself.  Several insights have been obtained over the past two decades, but it has remained an open problem to pinpoint the symmetry or other structure that causes this non-generic behavior.  Here we show that group theory provides the explanation: the governing equations for these systems arise naturally from the action of the group of conformal mappings of the unit disc to itself. This link unifies and explains the previous numerical and analytical results, and yields new constants of motion for this class of dynamical systems.}

\section{INTRODUCTION \label{INTRODUCTION}}

When a nonlinear system shows unexpectedly simple behavior, it may be a clue that some hidden structure awaits discovery.  

For example, recall the classic detective story~\cite{jackson90}  that began in the 1950s with the work of Fermi, Pasta, and Ulam~\cite{FPU, Weissert, zabu05}.  In their numerical simulations of a chain of anharmonic oscillators, Fermi et al. were surprised to find the chain returning almost perfectly, again and again, to its initial state.  The struggle to understand these recurrences led Zabusky and Kruskal~\cite{zabu65} to the discovery of solitons in the Korteweg--deVries equation, which in turn sparked a series of results showing that this equation possessed many conserved quantities---in fact, infinitely many~\cite{miura76}.  Then several other equations turned out to have the same properties.  At the time these results seemed almost miraculous.  But by the mid-1970s  the hidden structure responsible for all of them---the complete integrability of certain infinite-dimensional Hamiltonian systems~\cite{zakh71}---had been made manifest by the inverse scattering transform~\cite{gard67, ablo81} and Lax pairs~\cite{lax68}.  

Something similar, though far less profound, has been happening again in nonlinear science.  The broad topic is still coupled oscillators, but unlike the conservative oscillators studied by Fermi et al., the oscillators in question now are dissipative and have stable limit cycles.  This latest story began around 1990, when a few researchers noticed an enormous amount of neutral stability and seemingly low-dimensional behavior in their simulations of Josephson junction arrays---specifically, arrays of identical, overdamped junctions arranged in series and coupled through a common load~\cite{tsan91, tsan92, swift92, golo92, nich92}.  Then, just a year ago, Antonsen et al.~\cite{anto08} uncovered similarly low-dimensional dynamics in the periodically forced version of the Kuramoto model of biological oscillators~\cite{kura84, stro00, aceb05}.  This was particularly surprising because the oscillators in that model are non-identical.   

As in the soliton story, these numerical observations then inspired a series of theoretical advances.  These included the discovery of constants of motion~\cite{wata93, wata94}, and of a pair of transformations that established the low-dimensionality of the dynamics~\cite{wata93, wata94, goeb95, otta08, piko08, otta09}.  But what remained to be found was the final piece, the identification of the hidden structure.  Without it, it was unclear why the transformations and constants of motion should exist in the first place.  

In this paper we show that the group of M\"{o}bius transformations is the key to understanding this class of dynamical systems.  Our analysis unifies the previous treatments of Josephson arrays and the Kuramoto model, and clarifies the geometric and algebraic structures responsible for their low-dimensional behavior.  One spin-off of our approach is a new set of constants of motion; these generalize the constants found previously, and hold for a wider class of oscillator arrays.  

The paper is organized as follows.  To keep the treatment self-contained and to establish notation, Section~\ref{BACKGROUND} reviews the relevant background about coupled oscillators and the M\"{o}bius group.   In Section~\ref{MOBIUS GROUP REDUCTION} we show how to use M\"{o}bius transformations to reduce the dynamics of oscillator arrays with global sinusoidal coupling, a class that includes the Josephson and Kuramoto models as special cases.  The reduced flow lives on a set of invariant three-dimensional manifolds, arising naturally as the so-called group orbits of the M\"{o}bius group.  The results obtained in this way are then compared to previous findings (Section~\ref{CONNECTIONS TO PREVIOUS RESULTS}) and used  to generate new constants of motion via the classical cross ratio construction (Section~\ref{CHARACTERISTICS OF THE MOTION}).  We explore the dynamics on the invariant manifolds in Section~\ref{CHAOS IN JOSEPHSON ARRAYS}, and show that the phase portraits for resistively coupled Josephson arrays are filled with chaos and island chains, reminiscent of the pictures  encountered in Hamiltonian chaos and KAM theory.

\section{BACKGROUND \label{BACKGROUND}}

\subsection{Reducible systems with sinusoidal coupling\label{Reducible systems}}

The theory developed here was originally motivated by simulations of the governing equations for a series array of $N$ identical, overdamped Josephson junctions driven by a constant current and coupled through a resistive load.  As shown in Tsang et al.~\cite{tsan91}, the dimensionless circuit equations for this system can be written as
	\begin{equation} \label{jj_resistive}
	\dot{\phi_j} = \Omega-(b+1) \cos \phi_j + \frac{1}{N} \sum_{k=1}^N \cos \phi_k
	\end{equation}
for $j=1,\ldots, N$.  The physical interpretation need not concern us here; the important point for our purposes is that this set of $N$ ordinary differential equations (ODEs) displayed low-dimensional dynamics.  The same sort of low-dimensional behavior was later found in other kinds of oscillator arrays~\cite{golo92} as well as in Josephson arrays with other kinds of loads~\cite{tsan92, swift92, nich92}.  

Building on contributions from several teams of researchers~\cite{tsan91, tsan92, swift92, golo92, nich92}, Watanabe and Strogatz~\cite{wata94} showed that the system (\ref{jj_resistive}) could be reduced from $N$  ODEs to three ODEs, in the following sense.  Consider a time-dependent transformation from a set of constant angles $\theta_j$ to a set of functions $\phi_j(t)$, defined via
	\begin{equation} \label{WS_transformation}
	\tan\left[\frac{\phi_j(t)-\Phi(t)}{2}\right] = \sqrt{\frac{1+\gamma(t)}{1-\gamma(t)}} \tan\left[\frac{\theta_j-\Theta(t)}{2}\right]
	\end{equation}
for $j=1,\ldots, N$.   By direct substitution, one can check that the resulting functions $\phi_j(t)$ simultaneously satisfy all $N$ equations in (\ref{jj_resistive}) as long as the three variables $\Phi(t), \gamma(t)$ and $\Theta(t)$ satisfy a certain closed set of ODEs~\cite{wata94}. 

Watanabe and Strogatz also noted that the same transformation can be used to reduce any system of the form 
\begin{equation} \label{reducible}
	\dot{\phi_j} = f e^{i\phi_j} + g + \bar{f}e^{-i\phi_j} 
	\end{equation}
for $j = 1, \ldots, N$, where $f$ is any smooth, complex-valued, $2 \pi$-periodic function of the phases $\phi_1, \ldots, \phi_N$.  (Here the overbar denotes complex conjugate.  Also, note that $g$ has to be real-valued since $\dot{\phi_j}$ is real.)  The functions $f$ and $g$ are allowed to depend on time and on any other auxiliary state variables in the system, for example, the charge on a load capacitor or the current through a load resistor for certain Josephson junction arrays.  The key is that $f$ and $g$ must be the same for all oscillators, and thus do \emph{not} depend on the index $j$.  We call such systems \emph{sinusoidally coupled} because the dependence on $j$ occurs solely through the first harmonics $e^{i\phi_j}$ and $e^{-i\phi_j}$.

Soon after the transformation (\ref{WS_transformation}) was reported, Goebel~\cite{goeb95} observed that it could be related to fractional linear transformations, and he used this fact to simplify some of the calculations in Ref.~\cite{wata94}.  At that point, research on the reducibility of Josephson arrays paused for more than a decade.  The question of \emph{why}  this particular class of dynamical systems (\ref{reducible}) should be reducible by fractional linear transformations was not pursued at that time, but will be addressed in Section~\ref{MOBIUS GROUP REDUCTION}.

\subsection{Ott-Antonsen ansatz \label{Ott-Antonsen ansatz}}

Ott and Antonsen~\cite{otta08, otta09} recently reopened the issue of low-dimensional dynamics, with their discovery of an ansatz that collapses the infinite-dimensional Kuramoto model to a two-dimensional system of ODEs.  

To illustrate their ansatz in its simplest form, let us apply it to the class of identical oscillators governed by Eq.(\ref{reducible}), in the limit $N \rightarrow \infty$.  (Note that this step involves two simplifying assumptions, namely, that $N$ is infinitely large and that the oscillators are identical.  The Ott-Antonsen ansatz applies more generally to systems of non-identical oscillators with frequencies chosen at random from a prescribed probability distribution---indeed, this generalization was one of Ott and Antonsen's major advances---but it is not needed for the issues that we wish to address.)  In the limit $N \rightarrow \infty$, the evolution of the system (\ref{reducible}) is given by the continuity equation 
	\begin{equation} \label{continuity}
	\frac{\partial \rho}{\partial t} + \frac{\partial (\rho v)}{\partial \phi} = 0
	\end{equation}
where the phase density $\rho(\phi,t)$ is defined such that $\rho(\phi,t) \mathrm{d}\phi$ gives the fraction of phases that lie between $\phi$ and $\phi + \mathrm{d}\phi$ at time $t$, and where the velocity field is the Eulerian version of  (\ref{reducible}):
	\begin{equation} \label{velocity}
	v(\phi, t) = f e^{i\phi} + g + \bar{f}e^{-i\phi}.
	\end{equation}
Our earlier assumptions about the coefficient functions $f$ and $g$ now take the form that $f$ and $g$ may depend on $t$ but not on $\phi$.  The time-dependence of $f$ and $g$ can arise either explicitly (through external forcing, say) or implicitly (through the time-dependence of the harmonics of $\rho$ or any auxiliary state variables in the system).

Following Ott and Antonsen~\cite{otta08}, suppose $\rho$ is of the form
	\begin{equation} \label{OA_ansatz}
	\rho(\phi, t) = \frac{1}{2\pi} \biggl\lbrace 1+\sum_{n=1}^\infty \bigl(\bar{\alpha}(t)^ne^{in\phi}+\alpha(t)^ne^{-in\phi}\bigr) \biggr\rbrace
	\end{equation}
for some unknown function $\alpha$ that is independent of $\phi$. (Our definition of $\alpha$ is, however, slightly different from that in Ott and Antonsen~\cite{otta08}; our $\alpha$ is their $\bar \alpha$.)  Note that (\ref{OA_ansatz}) is just an algebraic rearrangement of the usual form for the Poisson kernel:
	\begin{equation} \label{Poisson}
	\rho(\phi) = \frac{1}{2\pi} \frac{1-r^2}{1-2r\cos(\phi-\Phi)+r^2}
	\end{equation}
where $r$ and $\Phi$ are defined via 
	\begin{equation} \label{alpha}
	\alpha = re^{i\Phi}.  
	\end{equation}
In geometrical terms, the ansatz (\ref{OA_ansatz}) defines a submanifold in the infinite-dimensional space of density functions $\rho$.  This \emph{Poisson submanifold} is two-dimensional and is parametrized by the complex number $\alpha$, or equivalently, by the polar coordinates $r$ and $\Phi$.

The intriguing fact discovered by Ott and Antonsen is that the Poisson submanifold is invariant: if the density is initially a Poisson kernel, it remains a Poisson kernel for all time.  To verify this, we substitute the velocity field (\ref{velocity}) and the ansatz (\ref{OA_ansatz}) into the continuity equation (\ref{continuity}), and find that the amplitude equations for each harmonic $e^{in\phi}$ are simultaneously satisfied if and only if $\alpha(t)$ evolves according to
	\begin{equation} \label{alpha_eqn}
	\dot{\alpha} = i\bigl(\bar{f}+g\alpha+f\alpha^2\bigr).
	\end{equation}

This equation can be recast in a more physically meaningful form in terms of the complex order parameter, denoted by $\langle z \rangle$ and defined as the centroid of the phases $\phi$ regarded as points $e^{i \phi}$ on the unit circle: 
	\begin{equation} \label{z}
	\langle z \rangle = \int_0^{2\pi} e^{i\phi} \rho(\phi, t) \mathrm{d}\phi.
	\end{equation}
By substituting (\ref{OA_ansatz}) into (\ref{z}) we find that $\langle z \rangle = \alpha$ for all states on the Poisson submanifold.  Hence, $\langle z \rangle$ satisfies the Riccati equation 
	\begin{equation} \label{riccati}
	\dot{\langle z \rangle} = i(\bar{f} + g\langle z \rangle + f \langle z \rangle^2).
	\end{equation}

When $f$ and $g$ are functions of $\langle z \rangle$ alone, as in mean-field models, Eq.(\ref{riccati}) constitutes a closed two-dimensional system for the flow on the Poisson submanifold.  More generally, the system will be closed whenever $f$ and $g$ depend on $\rho$ only through its Fourier coefficients.  We will show this explicitly in Subsection~\ref{Fourier Coefficients of the Phase Distribution}, by finding formulas for all the higher Fourier coefficients in terms of $\alpha$, and hence in terms of $\langle z \rangle$.  (However, as we will see, things become more complicated for states lying off the Poisson submanifold.  Then $\langle z \rangle$ no longer coincides with $\alpha$ and the closed system becomes three dimensional, involving $\psi$ as well as $\alpha$.)  

The work of Ott and Antonsen~\cite{otta08} raises several questions.  Why should the set of Poisson kernels be invariant?  What is the relationship, if any, between the ansatz (\ref{OA_ansatz}) and the transformation (\ref{WS_transformation}) studied earlier?  Why does  (\ref{WS_transformation}) reduce equations of the form (\ref{reducible}) to a three-dimensional flow, whereas (\ref{OA_ansatz}) reduces them to a two-dimensional flow?  

As we shall see, the answers have to do with the properties of the group of conformal mappings of the unit disc to itself.  Before showing how this group arises naturally in the dynamics of sinusoidally coupled oscillators, let us recall some of its relevant properties.  

\subsection{M\"{o}bius group \label{Mobius group}}

Consider the set of all fractional linear transformations $F: \mathbb{C} \rightarrow \mathbb{C}$ of the form 
	\begin{equation} \label{FLT}
	F(z) = \frac{a z + b}{c z + d},
	\end{equation}
where $a,b,c$ and $d$ are complex numbers, and the numerator is not a multiple of the denominator (that is, $ad-bc \neq 0$).  This family of functions carries the structure of a group.  The group operation is composition of functions, the identity element is the identity map, and inverses are given by inverse functions.    

Of most importance  to us is a subgroup $G$---which we refer to as the \emph{M\"{o}bius group}---consisting of those fractional linear transformations that map the open unit disc  
$\mathbb{D} = \{z \in \mathbb{C}: |z| < 1\}$  onto itself in a one-to-one way. These transformations and their inverses are analytic on $\mathbb{D}$ and map its boundary (the unit circle $S^1 = \{z \in \mathbb{C}: |z| = 1\}$) to itself. All such automorphisms of the disc can be written~\cite{stein} in the form 
	\begin{equation} \label{usual_automorphism_parametrization}
	F(z) = e^{i\varphi} \frac{\alpha -z}{1- \bar{\alpha} z},
	\end{equation}
for some $\varphi \in \mathbb{R}$ and $\alpha \in \mathbb{D}$.  The M\"{o}bius group $G$ is in fact a three-dimensional Lie group, with real parameters $\varphi,$ Re$(\alpha)$, and Im$(\alpha)$.  

However, it turns out that a different parametrization of $G$ will be more notationally convenient in what follows, in the sense that it simplifies comparisons between our results and those in the prior literature.  Specifically, we will view a typical element of $G$ as a mapping $M$ from the unit disc in the complex $w$-plane to the unit disc in the complex $z$-plane, with parametrization given by 
	\begin{equation} \label{our_automorphism_parametrization}
	z=M(w) = \frac{e^{i\psi}w + \alpha}{1 + \bar{\alpha}e^{i\psi}w}
	\end{equation}
where $\alpha \in \mathbb{D}$ and $\psi \in \mathbb{R}$.  Note that the inverse mapping 
	\begin{equation} \label{inverse_parametrization}
	w=M^{-1}(z) = e^{-i\psi} \frac{z - \alpha}{1 - \bar{\alpha} z}
	\end{equation}
has an appearance closer to that of the standard parametrization (\ref{usual_automorphism_parametrization}).  

A word about terminology: our definition of the M\"{o}bius group is not the conventional one.  Usually this term denotes the larger group of all fractional linear transformations (or bilinear transformations, or linear fractional transformations), whereas we reserve the adjective M\"{o}bius for the subgroup $G$ and its elements.  Thus, from now on, when we say \emph{M\"{o}bius transformation} we specifically mean an element of the subgroup $G$ consisting of analytic automorphisms of the unit disc.

\section{M\"{O}BIUS GROUP REDUCTION \label{MOBIUS GROUP REDUCTION}}

In this section we show that if the equations for the oscillator array are of the form (\ref{reducible}), then the oscillators' phases $\phi_j(t)$ evolve according to the action of the M\"{o}bius group on the complex unit circle:  
\begin{equation} \label{groupaction}
e^{i\phi_j(t)} = M_t(e^{i\theta_j}), 
\end{equation}
for $j=1, \ldots, N$, where $M_t$ is a one-parameter family of M\"{o}bius transformations and 
$\theta_j$ is a constant (time-independent) angle.  
In other words, the time-$t$ flow map for the system is always a M\"{o}bius map.  

Incidentally, this result is consistent with a basic topological fact: we know that different oscillators cannot pass through each other on $S^1$ under the flow, so we expect the time-$t$ flow map to be an orientation-preserving homeomorphism of $S^1$ onto itself---and indeed any M\"{o}bius map is.

We begin the analysis with an algebraic method similar to that in Goebel~\cite{goeb95}.   Then, in Sections~\ref{Geometric Method of Finding dotalpha} and~\ref{Geometric Method of Finding dotpsi}, we adopt a geometrical perspective and show that it answers several questions left open by the first method.

\subsection{Algebraic Method \label{Algebraic Method}}

Parametrize the one-parameter family of M\"{o}bius transformations as
	\begin{equation} \label{2}
	M_t(w) = \frac{e^{i\psi}w + \alpha}{1 + \bar{\alpha}e^{i\psi}w}
	\end{equation}
where $|\alpha(t)| < 1$ and $\psi(t) \in \mathbb{R}$, and let 
	\begin{equation} \label{w-defn}
	w_j =  e^{i \theta_j}. 
	\end{equation}
To verify that (\ref{2}) gives an exact solution of (\ref{reducible})---subject to the constraint that the M\"{o}bius parameters $\alpha(t)$ and $\psi(t)$ obey appropriate ODEs, to be determined---we compute the time-derivative of $\phi_j(t) = -i \log M_t(w_j)$, keeping in mind that $w_j$ is constant:
	\begin{equation} \label{3}
	\dot{\phi_j} = \frac{\dot{\psi}e^{i\psi}w_j - i \dot{\alpha}}{e^{i\psi}w_j + \alpha} + \frac{(i\dot{\bar{\alpha}}
		- \bar{\alpha}\dot{\psi})e^{i\psi}w_j}{1 + \bar{\alpha}e^{i\psi}w_j}.
	\end{equation}
From (\ref{inverse_parametrization}), we get
	\begin{equation} \label{4}
	e^{i\psi}w_j = \frac{e^{i\phi_j} - \alpha}{1 - \bar{\alpha}e^{i\phi_j}}
	\end{equation}
which when substituted into (\ref{3}) yields
	\begin{equation} \label{5}
	\dot{\phi_j} = Re^{i\phi_j} + \frac{\dot{\psi} + i\bar{\alpha}\dot{\alpha} - \alpha(i\dot{\bar{\alpha}} - \bar{\alpha}\dot{\psi})}{1 - |\alpha|^2} + \bar{R}e^{-i\phi_j}
	\end{equation}
where $R = (i\dot{\bar{\alpha}} - \bar{\alpha}\dot{\psi})/(1 - |\alpha|^2)$.  

Note that Eq.(\ref{5}) falls precisely into the algebraic form required by (\ref{reducible}).  Thus, to derive the desired ODEs for $\alpha(t)$ and $\psi(t)$, we now subtract (\ref{5}) from (\ref{reducible}) to obtain $N$ equations of the form $0 = C_1 e^{i\phi_j} + C_0 + C_{-1} e^{-i\phi_j}$, for $j=1, \ldots, N$.  If the system contains at least three distinct oscillator phases, then $C_1$, $C_0$, and $C_{-1}$ must generically be zero.  Explicitly,
	\begin{equation} \label{6}
	f = \frac{i\dot{\bar{\alpha}} - \bar{\alpha}\dot{\psi}}{1 - |\alpha|^2}, \;\;\;
	g = \frac{\dot{\psi} + i\bar{\alpha}\dot{\alpha} - \alpha(i\dot{\bar{\alpha}} - \bar{\alpha}\dot{\psi})}{1 - |\alpha|^2}.
	\end{equation}
The system (\ref{6}) can be algebraically rearranged to give
	\begin{subequations} \label{7}
	\begin{align}
	\dot{\alpha} &= i(f\alpha^2 + g\alpha + \bar{f})       \label{7a} \\
	\dot{\psi}   &= f\alpha + g + \bar{f}\bar{\alpha}. \label{7b}	
	\end{align}
	\end{subequations}
	
Equations (\ref{7a}) and (\ref{7b}) have been derived previously; they appear as Eqs.(10) and (11), respectively, in Pikovsky and Rosenblum's work~\cite{piko08}, where they were derived by applying the transformation (\ref{WS_transformation}).  Both their approach and the one above are certainly quick and clean, but they require us to guess the transformation ahead of time, and reveal little about why this transformation works. 

Incidentally, observe that under the change of variables $z_j~=~e^{i\phi_j}$, (\ref{reducible}) becomes
	\begin{equation} \label{1.1}
	\dot{z_j} = i(f z_j^2 + g z_j + \bar{f}).
	\end{equation}
Equation~(\ref{1.1}) is a Riccati equation with the form of (\ref{7a})---another coincidence that seems a bit surprising when approached this way.  In the following subsection, we will see how these Ricatti equations emerge naturally from the infinitesimal generators of the M\"{o}bius group.

\subsection{Geometric Method of Finding $\dot{\alpha}$ \label{Geometric Method of Finding dotalpha}}

Now we change our view of M\"{o}bius maps slightly.  Instead of thinking of $M$ as a map from the $w$-plane to the $z$-plane, we view it as a map from the $z$-plane to itself.  This requires a small and temporary change in notation, but it makes things clearer, especially when we start to discuss differential equations on the complex plane.     

We begin by recalling some basic facts and definitions.  Suppose the coupled oscillator system contains just three distinct phases among its $N$ oscillators.  Then by a property of M\"{o}bius transformations, there exists a unique M\"{o}bius transformation from any point $\bm{z}_1 = (e^{i\theta_1}, e^{i\theta_2}, e^{i\theta_3})$ to any other point $\bm{z}_2 = (e^{i\phi_1}, e^{i\phi_2}, e^{i\phi_3})$ in the state space $S^1~\times~S^1~\times~S^1$.  If the system instead contains only one or two distinct phases, many M\"{o}bius transformations will take $\bm{z}_1$ to $\bm{z}_2$, so we can still reach every point of the phase space from every other point.  However, if the system contains more than three distinct phases, say $N$, then there is not in general a M\"{o}bius transformation that transforms $\bm{z}_1 = (e^{i\theta_1}, e^{i\theta_2}, e^{i\theta_3}, \dotsc, e^{i\theta_N})$ to $\bm{z}_2 = (e^{i\phi_1}, e^{i\phi_2}, e^{i\phi_3}, \dotsc, e^{i\phi_N})$; only some points are accessible from $\bm{z}_1$, while others are not.

In the language of group theory, we say that $\bm{z}_2$ is in the \textit{group orbit} of $\bm{z}_1$ if there exists a M\"{o}bius map $M$ such that $\bm{z}_2 = M(\bm{z}_1)$.  Then, as a direct consequence of the fact that M\"{o}bius maps form a three-parameter group $G$ under composition, the group orbits of $G$ partition the phase space into three-dimensional manifolds (when the phase space is at least three-dimensional).

To compute infinitesimal generators for $G$, we compute the time derivatives of the three one-parameter families of curves corresponding to the three parameters of $G$:  $\psi$, $\text{Re} (\alpha)$ and $\text{Im} (\alpha)$.  Each of the three families is obtained from the M\"{o}bius transformation by setting two of the three parameters to zero, and leaving the remaining parameter free.  For example, if we set $t = 0$ at $\bm{z} = (z_1, \dotsc, z_N)$, these three families are
	\begin{equation} \label{7.1}
	\begin{split}
	M_1(\bm{z}) &= e^{it}\bm{z}                    \\
	M_2(\bm{z}) &= \frac{\bm{z} - t}{1 - t\bm{z}}  \\
	M_3(\bm{z}) &= \frac{\bm{z} + it}{1 - it\bm{z}}
	\end{split}
	\end{equation}
where $M_1(\bm{z})$ is written in place of $(M_1(z_1), \dotsc, M_1(z_N))$ and likewise for $M_2(\bm{z})$ and $M_3(\bm{z})$.  We continue using this shorthand in subsequent equations, writing $h(\bm{z})$ in place of $(h(z_1), \dotsc, h(z_N))$ for any one-parameter function $h$.

The time derivatives of the curves in (\ref{7.1}) evaluated at $t = 0$ then give a set of infinitesimal generators for $G$:
	\begin{equation} \label{7.2}
	\begin{split}
	\bm{v}_1 &= i\bm{z}      \\
	\bm{v}_2 &= \bm{z}^2 - 1 \\
	\bm{v}_3 &= i\bm{z}^2 + i.
	\end{split}
	\end{equation}
Note that these three generators point out into the full $N$-dimensional complex space $\mathbb{C}^N$, as expected.

Meanwhile, if we substitute $f = -ih_1 + h_2$ (where $h_1$ and $h_2$ are real functions) into the original Riccati dynamics (\ref{1.1}), we can rewrite this equation of motion in terms of the three infinitesimal generators:
	\begin{equation} \label{7.3}
	\dot{\bm{z}} = i\bm{z}g + (\bm{z}^2 - 1)h_1 + (i\bm{z}^2 + i)h_2.
	\end{equation}

The implication of the rewritten form (\ref{7.3}) is then given by a theorem from Lie theory:  if $L$ is a Lie group acting on a submanifold with linearly independent infinitesimal generators $\bm{v}_1, \dotsc, \bm{v}_n$, and $\bm{v}$ is a vector field of the form $\bm{v} = c_1\bm{v}_1 + \dotsb + c_n\bm{v}_n$ where the coefficients $c_k$ depend only on time $t$, then the trajectory of the dynamics $\dot{\bm{z}} = \bm{v}$ from any initial condition $\bm{z}_0$ can be expressed in the form $\{A_t(\bm{z}_0)\}$ for a unique family $\{A_t\} \subset L$ parameterized by $t$.  Since the M\"{o}bius group is a complex Lie group, this result can be applied directly to conclude (\ref{7.3}) has the solution $\bm{z}(t) = M_t(\bm{z}_0)$ where $\{M_t\}$ is a unique one-parameter family of M\"{o}bius transformations.

Although we have so far assumed that the components $z_k$ of $\bm{z}$ lie on the complex unit circle, both (\ref{2}) and (\ref{7.3}) extend naturally to all of $\mathbb{C}^N$.  This implies that $z_0 = 0$ must evolve as $z(t) = M_t(0)$ for some family $\{M_t\}$.  However, Eq.~(\ref{2}) shows that $M(0) = \alpha$ for all $M \in G$.  So $z(t) = M_t(0) = \alpha$ for all $t$, meaning that $\alpha(t)$ satisfies (\ref{7.3}).  Since (\ref{7.3}) is just a rewriting of (\ref{1.1}), the dynamics (\ref{7a}) for $\alpha$ that we derived earlier are now placed in a geometrical context.  This approach reveals that $\alpha(t)$ is just the image of the origin under a one-parameter family of M\"{o}bius maps, applied to any one complex plane of $\mathbb{C}^N$. 

It is even more illuminating to compute the infinitesimal generators within the $N$-fold torus $\mathbb{T}^N$ of phase values, i.e., the quantities $\bm{u}_k = -i \frac{d}{dt} \log M_k(e^{i\bm{\phi}})|_{t = 0}$.  These turn out to be
	\begin{equation} \label{7.4}
	\begin{split}
	\bm{u}_1 &= (1, \dotsc, 1)   \\
	\bm{u}_2 &= 2 \sin \bm{\phi} \\
	\bm{u}_3 &= 2 \cos \bm{\phi}.
	\end{split}
	\end{equation}
When expressed in terms of these infinitesimal generators, the equation of motion (\ref{7.3}) becomes
	\begin{equation} \label{7.4.1}
	\dot{\bm{\phi}} = g + (2 \sin \bm{\phi})h_1 + (2 \cos \bm{\phi})h_2
	\end{equation}
which is precisely what we earlier referred to as a sinusoidally coupled system (\ref{reducible}), and whose solution must therefore be of the form $\bm{\phi}_t =  -i\log M_t (e^{i\bm{\theta}})$ for some $M_t \in G$.  

This calculation finally clarifies what is so special about sinusoidally coupled systems (\ref{reducible}): they are induced naturally by a flow on the M\"{o}bius group.  This fact underlies their reducibility and all their other beautiful (but non-generic) properties.

\subsection{Geometric Method of Finding $\dot{\psi}$ \label{Geometric Method of Finding dotpsi}}

We turn next to the dynamics of $\psi$.  As we will show in the next section, the action of the M\"{o}bius transformation involves a clockwise rotation of the oscillator phase density $\rho(\phi,t)$ by $\arg(\alpha) - \psi$ and a counterclockwise rotation by $\arg(\alpha)$.  Hence, $\psi(t)$ may be viewed as the overall counterclockwise rotation of the distribution at time $t$ relative to the initial distribution at $t = 0$.  

To support this interpretation, we show here that $\dot{\psi}$ equals the average value of the vector field on the circle, given by 
	\begin{equation} \label{averagephidot}
	\langle \dot{\phi} \rangle = \frac{1}{2\pi} \int_{S^1} \dot{\phi} \, d\theta.
	\end{equation}
Observe the right side of the integrand (\ref{3}) has two terms:
	\begin{equation} \label{7.4.2}
	\begin{split}
	R_1(w) &= \frac{\dot{\psi}e^{i\psi}w - i \dot{\alpha}}{e^{i\psi}w + \alpha} \\
	R_2(w) &= \frac{(i\dot{\bar{\alpha}} - \bar{\alpha}\dot{\psi})e^{i\psi}w}{1 + \bar{\alpha}e^{i\psi}w}.
	\end{split}
	\end{equation}
By Cauchy's formula,
	\begin{equation} \label{7.5}
	\frac{1}{2\pi i} \int_{S^1} R_2(w) \frac{dw}{w} = R_2(0) = 0.
	\end{equation}
So $\langle \dot{\phi} \rangle$ simplifies to
	\begin{equation} \label{7.6}
	\langle \dot{\phi} \rangle = \frac{1}{2\pi i} \int_{S^1} R_1(w) \frac{dw}{w}.
	\end{equation}
Note that $R_1(w)$ has a pole in the unit disc, so we make the change of variables $w \rightarrow w^{-1}$ to move this pole outside the circle.  Evaluating the resulting integral yields 
	\begin{equation} \label{7.7}
	\frac{1}{2\pi i} \! \int_{S^1} \! R_1(w) \frac{dw}{w} 
		= \frac{1}{2\pi i} \! \int_{S^1} \! \frac{\dot{\psi} - i\dot{\alpha} e^{-i\psi}w}{1 + \alpha e^{-i\psi}w} \frac{dw}{w} = \dot{\psi}
	\end{equation}
which completes the demonstration that $\langle \dot{\phi} \rangle = \dot{\psi}$.

We can now go back and evaluate the average vector field in a different way to find the differential equation that governs $\psi(t)$.  Differentiating $\phi = -i \log M_t(w)$ with respect to time and substituting the result into $\dot{\psi} = \frac{1}{2\pi} \int_{S^1} \dot{\phi} \, d\theta$, we obtain
	\begin{equation} \label{7.8}
	\dot{\psi} = \frac{1}{2\pi i} \int_{S^1} \frac{\dot{M}_t(w)}{M_t(w)} \frac{dw}{iw}.
	\end{equation}
Since $M_t$ obeys the Ricatti equation, we can eliminate $\dot{M}_t$ in the numerator above to get
	\begin{equation} \label{7.9}
	\dot{\psi} = \frac{1}{2\pi i} \int_{S^1} (f M_t(w) + g + \bar{f} M_t(w)^{-1}) \frac{dw}{w}.
	\end{equation}
There are three integrals to evaluate here.   The third one involves a term $M_t(w)^{-1}$ which has a pole inside the unit circle, so we do the same change of variables as before, $w \rightarrow w^{-1}$, to move the pole outside.  The corresponding integral then simplifies to
	\begin{equation} \label{7.10}
	\frac{1}{2\pi i} \! \int_{S^1} \! \! \! M_t(w)^{-1} \frac{dw}{w} 
		= \frac{1}{2\pi i} \! \int_{S^1} \! \frac{e^{-i\psi}w + \bar{\alpha}}{1 + \alpha e^{-i\psi}w} \frac{dw}{w} = \bar{\alpha}
	\end{equation}
where the final integration follows from Cauchy's formula.  Similiarly, we use Cauchy's formula to integrate the first and second terms of the integrand in (\ref{7.9}), and thereby obtain the desired differential equation for $\psi$, thus rederiving (\ref{7b}) found earlier.

\section{CONNECTIONS TO PREVIOUS RESULTS \label{CONNECTIONS TO PREVIOUS RESULTS}}

\subsection{Relation to the Watanabe-Strogatz Transformation \label{Relation to the Watanabe-Strogatz Transformation}}

It is natural to ask how the trigonometric transformation (\ref{WS_transformation}) used in earlier studies~\cite{wata93, wata94, piko08} relates to the M\"{o}bius transformation (\ref{2}) used above.  As we will see, (\ref{WS_transformation}) may be viewed as a restriction of (\ref{2}) to the complex unit circle.  

First, by trigonometric identities, we have
	\begin{equation} \label{9}
	\tan\left[\frac{\phi-\Phi}{2}\right] = i\frac{1- e^{i(\phi-\Phi)}}{1 + e^{i(\phi-\Phi)}}.
	\end{equation}
To connect this to M\"{o}bius transformations, consider what happens when we apply the  map defined by (\ref{2}) to a point $w = e^{i\theta}$ on the unit circle.  Since the image is also a point on the unit circle, it can be written as $M(e^{i\theta}) = e^{i\phi}$ for some angle $\phi$.  Next let $\alpha = re^{i\Phi}$ and divide both sides of (\ref{2}) by $e^{i\Phi}$.  Thus 
	\begin{equation} \label{trigtoMobius}
	e^{i(\phi - \Phi)} = \frac{e^{i(\theta-\Theta)} + r}{1 + re^{i(\theta-\Theta)}}
	\end{equation}
where $\Theta = \Phi - \psi$.  Substitution of (\ref{trigtoMobius}) into the right side of (\ref{9}) gives
	\begin{equation} \label{10}
	\tan\left[\frac{\phi-\Phi}{2}\right] = \frac{1-r}{1+r} \bigg(i\frac{1- e^{i(\theta-\Theta)}}{1 + e^{i(\theta-\Theta)}}\bigg).
	\end{equation}
By the identity (\ref{9}), Eq.(\ref{10}) is equivalent to (\ref{WS_transformation}) with $\gamma = -2r/(1+r^2)$.  

We can now see how the M\"{o}bius parameters $\alpha$ and $\psi$ operate on the set of $e^{i\theta}$ in $\mathbb{C}$.  From the relationships between $\Theta$, $\gamma$, $\Phi$ and the M\"{o}bius parameters, the initial phase density is first rotated clockwise around $S^1$ by $\arg(\alpha) - \psi$, then squeezed toward one side of the circle as a function of $|\alpha|$, and afterwards rotated counterclockwise by $\arg(\alpha)$.  The squeeze, which takes uniform distributions to Poisson kernels, can be thought of as a composition of inversions, dilations and translations in the complex plane.

\subsection{Invariant Manifold of Poisson Kernels \label{Invariant Manifold of Poisson Kernels}}

In Section~\ref{Ott-Antonsen ansatz}  and in a previous paper~\cite{marv09}, we used the Ott-Antonsen ansatz (\ref{OA_ansatz}) to show that systems of identical oscillators with global sinusoidal coupling contain a degenerate two-dimensional manifold among the three-dimensional leaves of their phase space foliation.  This two-dimensional manifold, which we called the Poisson submanifold, consists of phase densities $\rho(\phi,t)$ that have the form of a Poisson kernel.  We now rederive these results within the framework of M\"{o}bius transformations.  

Let $T$ denote one instance of the transformation (\ref{WS_transformation}); in other words, fix the parameters $\Phi$, $\gamma$ and $\Theta$ and let $\phi = T(\theta)$.  Let $\mu$ denote the normalized uniform measure on $S^1$; thus 
	\begin{equation} \label{uniform_measure}
	d\mu(\theta) = \frac{1}{2\pi} d\theta.  
	\end{equation}
The transformation $T$ maps $\mu$ to the measure $T_*\mu$, and, by the usual formula for transformation of single-variable measures, we have $d(T_*\mu)(\phi) = \frac{1}{2\pi} T^{-1}(\phi)'d\phi$, where the prime denotes differentiation by $\phi$.  From this equation it follows that $d(T_*\mu)(\phi)$ has the form of the Poisson kernel, because the inverse of the M\"{o}bius transformation (\ref{2}) is
	\begin{equation} \label{11}
	M^{-1}(z) = e^{-i\psi} \frac{z - \alpha}{1 - \bar{\alpha}z}
	\end{equation}
which implies
	\begin{equation} \label{12}
	T^{-1}(\phi) = -\psi - i\log(e^{i\phi} - \alpha) + i\log(1 - \bar{\alpha}e^{i\phi}).
	\end{equation}
Then by differentiation and algebraic rearrangement, we obtain
	\begin{equation} \label{13}
	T^{-1}(\phi)' = \frac{1 - r^2}{1 - 2r\cos(\phi-\Phi) + r^2}.
	\end{equation}
The integral of $T^{-1}(\phi)'$ over $[0,2\pi)$ is $2\pi$, so $d(T_*\mu)(\phi)$ is indeed a normalized Poisson kernel.

Finally, if the phase distribution $d(T_*\mu)(\phi)/d\phi$ ever takes the form of a Poisson kernel with parameters $r = r_0$ and $\Phi = \Phi_0$, then we can set $r(0) = r_0$, $\Phi(0) = \Phi_0$ and $d\mu(\theta) = \frac{1}{2\pi}d\theta$, and the above calculation shows that $d(T_*\mu)(\phi)/d\phi$ remains a Poisson kernel for all future and past times.  Hence, the set of normalized Poisson kernels constitutes an invariant submanifold of the infinite-dimensional phase space.

The above demonstration also reveals that the Poisson submanifold has dimension $k + 2$ where $k$ is the number of state variables besides $\alpha$, $\psi$ and the oscillator phases.  More concretely, it implies that when the system lies on the Poisson submanifold, we can write $\dot{\alpha}$ as depending only on $\alpha$; it is not possible to require $\dot{\alpha}$ to depend on $\psi$ in any real coupling scheme.

To see this, we first consider the case in which the system is closed and there are no additional state variables.  Suppose $\dot{\alpha}$ does \textit{not} depend only on $\alpha$.  Then some of the state space trajectories cross when projected onto the unit disc of $\alpha$ values.  At the point of any crossing, the phase density $\rho(\phi,t)$ has multiple $\dot{\alpha}$ values.  But by (\ref{13}), the phase density depends only on $\alpha$, so there is nothing in the state space that can distinguish between the different $\dot{\alpha}$ values at that point.  Hence, $\dot{\alpha}$ must be expressible in terms of $\alpha$ alone.  By an analogous argument, $\dot{\alpha}$ is also independent of $\psi$ on the Poisson submanifold when there are $k$ other state variables besides the oscillator phases and M\"{o}bius parameters.

On the other hand, if the time-dependence of $\dot{\alpha}$ arises only via a dependence on $\alpha$, then $r$ and $\Phi$ decouple from $\psi$ and the dynamics are two-dimensional regardless of whether the system is evolving on the Poisson submanifold or not.  Observe that we can always force $\psi$-independence for $\dot{\alpha}$ by throwing away enough information about the locations of the other phases.  For instance, in the extreme, we may simply make $f$ and $g$ constant.  

Finally, even when $\dot{\alpha}$ does not depend solely on $\alpha$, the dynamics still may be two dimensional.  For example, in the case of completely integrable systems~\cite{wata93}, the variables $r$ and $\Phi - \psi$ decouple from $\Phi$ to foliate the phase space with two-dimensional tori.

\section{CHARACTERISTICS OF THE MOTION \label{CHARACTERISTICS OF THE MOTION}}

\subsection{Cross Ratios as Constants of Motion \label{Cross Ratios as Constants of Motion}}

The reduction of (\ref{reducible}) by the three-parameter M\"{o}bius group suggests that the corresponding system of coupled oscillators should have $N - 3$ constants of motion.  As we will see, these conserved quantities are given by the cross ratios of the points $z_j = e^{i\phi_j}$ on $S^1$.  Recall from complex analysis~\cite{conway} that the \emph{cross ratio} of four distinct points $z_1, z_2, z_3, z_4 \in \mathbb{C} \cup \{\infty\}$ is 
	\begin{equation} \label{14.1}
	(z_1, z_2, z_3, z_4) = \frac{z_1 - z_3}{z_1 - z_4} \cdot \frac{z_2 - z_4}{z_2 - z_3}
	\end{equation}
This quantity is conserved under M\"{o}bius transformations:  for all $\alpha$ and $\psi$, $(M(z_1), M(z_2), M(z_3), M(z_4)) = (z_1, z_2, z_3, z_4)$.  Hence, the $N!/(N-4)!$ cross ratios of the $N$ oscillator phases remain constant along the trajectories in phase space.  We denote the constant value of $(z_1, z_2, z_3, z_4)$ as $\lambda_{1234}$.  Of course, we could have defined the cross ratio for four-tuples of \textit{non-distinct} points as well, but these quantities are trivially conserved regardless of the dynamics and hence do not reduce the dimension of the phase space.

To show that exactly $N - 3$ of the cross ratios are independent, consider the sequence:  $\{(z_1, z_2, z_3, z_4),$ $(z_2, z_3, z_4, z_5), \dotsc , (z_{N-3}, z_{N-2}, z_{N-1}, z_N)\}$.  Each cross ratio in the sequence includes a new point not in the cross ratios preceding it and therefore must be independent of them.  Hence, there are at least $N - 3$ independent cross ratios.  With a bit more work (see the Appendix), we can also confirm that the rest of the cross ratios are functionally dependent on these $N - 3$ integrals.  

Since the state space of the phases is an $N$-fold torus of real variables, we expect that each of the constants of motion can be expressed in terms of real functions and variables.  Indeed, if $z_1$, $z_2$, $z_3$, $z_4$ lie on the unit circle, then the cross ratio $(z_1, z_2, z_3, z_4)$ lies on $\mathbb{R} \cup \{\infty\}$.  We see this explicitly by pulling out $e^{\frac{i}{2}(\phi_1 + \phi_3)}$ from the factor $(e^{i\phi_1} - e^{i\phi_3})$ of $(e^{i\phi_1}, e^{i\phi_2}, e^{i\phi_3}, e^{i\phi_4})$, and likewise for the other three factors, and then canceling the factors $e^{\frac{i}{2}(\phi_1 + \phi_2 + \phi_3 + \phi_4)}$ in the numerator and denominator to find
	\begin{equation} \label{17.1}
	(e^{i\phi_1}, e^{i\phi_2}, e^{i\phi_3}, e^{i\phi_4}) = \frac{S_{13} S_{24}}{S_{14} S_{23}}
	\end{equation}
where 
	\begin{equation} \label{S_ij}
	S_{ij} = \sin\left[\frac{\phi_i - \phi_j}{2}\right].  
	\end{equation}

This way of writing the cross ratio also suggests a relationship with the constants of motion reported by Watanabe and Strogatz~\cite{wata93, wata94} for completely integrable systems (those with $f = \frac{1}{2}e^{i\delta}\overline{\langle z \rangle}$ and $g = 0$, where $\langle z \rangle$ is the phase centroid (\ref{z})).  These constants of motion, which we will call \textit{WS integrals}, take the form
	\begin{equation} \label{17.2}
	I = S_{12} S_{23} \dotsm S_{(N-1)N} S_{N1}
	\end{equation}
where any permutation of the indices generates another WS integral.  As previously demonstrated~\cite{wata94}, exactly $N - 2$ of the $N!$ index permutations of (\ref{17.2}) are functionally independent.  

As we might anticipate, the WS integrals imply that the cross ratios are constants of motion:  consider two distinct WS integrals $I = S_{ik}S_{kl}S_{lj}\Pi$ and $I' = S_{il}S_{lk}S_{kj}\Pi$, where $\Pi$ denotes the remaining product of factors.  Assume $\Pi$ is the same for both $I$ and $I'$.  Then $I/I' = -\lambda_{ijkl}$.  Since $i$, $j$, $k$, $l$ are arbitrary, we can generate all cross ratios via this procedure.

Additionally, if a single WS integral holds for a system in which the cross ratios are invariant, then \textit{all} WS integrals hold, since we can arbitrarily permute the indices of the first WS integral by sequences of transpositions of the form $I = -\lambda_{ijkl} I'$ in which $l$ and $k$ are interchanged.

\subsection{Fourier Coefficients of the Phase Distribution \label{Fourier Coefficients of the Phase Distribution}}

When we introduced $f$ and $g$ in Section~\ref{BACKGROUND}, we required that they depend on the phases \emph{only} through the Fourier coefficients of the phase density $\rho(\phi,t)$.  Since the centroid (\ref{z}) is the Fourier coefficient corresponding to the first harmonic $e^{-i\phi}$, this condition is met by standard Kuramoto models, Josephson junction series arrays, laser arrays and many other well-studied systems of globally coupled oscillators.  

Our goal now is to show that this condition implies the closure of (\ref{7}), in the sense that $\dot{\alpha}$ and $\dot{\psi}$ depend only on $\alpha$ and $\psi$.  To do so, we will show that the Fourier coefficient of all higher harmonics $e^{-im\phi}$ for any integer $m$ may be expressed in terms of $\alpha$ and $\psi$.

For a fixed measure $\mu(\theta)$ on $[0,2\pi)$ and a transformation $T(\theta) = -i \log M(\theta)$ of this measure via the M\"{o}bius map $M$, the Fourier coefficient of $e^{-im\phi}$ is given by
	\begin{equation} \label{18}
	\langle z^m \rangle = \int_{S^1} e^{im\phi} d(T_*\mu)(\phi) = \int_{S^1} M(e^{i\theta})^m d\mu(\theta).
	\end{equation}
We use the notation $\langle z^m \rangle$ as a reminder that $\langle z \rangle$ is the phase centroid.  

We assume that we can take a Fourier expansion of $\mu(\theta)$, so 
	\begin{equation} \label{Fourier_expansion_of_measure}
	d\mu(\theta) = \frac{1}{2\pi} \sum_{n = -\infty}^\infty c_n e^{in\theta} d\theta
	\end{equation}
where the constants $c_n$ are independent of $\theta$.  Since the phase distribution must be real and normalized, we know that $c_{-n} = \bar{c}_n$ and $c_0 = 1$, so we can write
	\begin{equation} \label{19}
	d\mu = \frac{1}{2\pi i} \biggl(1 + P(w) + \overline{P(w)}\biggr) \frac{dw}{w}
	\end{equation}
where $w = e^{i\theta}$ and $P(w) = \sum_{n=1}^\infty c_n w^n$.  The formula for $\langle z^m \rangle$ then becomes:
	\begin{equation} \label{20}
	\langle z^m \rangle = \frac{1}{2\pi i} \int_{S^1} M(w)^m \biggl(1 + P(w) + \overline{P(w)}\biggr) \frac{dw}{w}.
	\end{equation}
Now, $M(w)^m (1 + P(w))$ is analytic on the open disc $\mathbb{D}$ and $M(0)^m (1 + P(0)) = \alpha^m$.  Meanwhile, the remaining term of the integrand of (\ref{20}) has the complex conjugate
	\begin{equation} \label{22}
	\frac{\overline{M(w)}^m P(w)}{w} = \biggl(\frac{1 + \bar{\alpha}e^{i\psi}w}{e^{i\psi}w + \alpha}\biggr)^m \frac{P(w)}{w}
	\end{equation}
which features an order-1 pole at $w = 0$ and an order-$m$ pole at $w = -e^{-i\psi}\alpha$.  The first residue evaluates to zero, while the second is given by
	\begin{equation} \label{24}
	\frac{e^{-im\psi}}{(m-1)!}\frac{d^{m-1}}{dw^{m-1}}\biggl[(1 + \bar{\alpha}e^{i\psi}w)^m \frac{P(w)}{w}\biggr] \biggl|_{w = -e^{-i\psi}\alpha.}
	\end{equation}
Therefore, $\langle z^m \rangle$ is equal to $\alpha^m$ added to the complex conjugate of this second residue:
	\begin{multline} \label{24.5}
	\langle z^m \rangle = \alpha^m + \sum_{k=0}^{m-1} \frac{(1 - |\alpha|^2)^{k+1}}{k!} \\
		\times \sum_{n=0}^{\infty} (-1)^n \frac{(n+k)!}{n!} \bar{c}_{n+k+1} e^{i(m+n)\psi} \bar{\alpha}^n.
	\end{multline}
For example, the centroid may be written in terms of $\alpha$ and $\psi$ as
	\begin{equation} \label{25}
	\langle z \rangle = \alpha + (|\alpha|^2 - 1) \sum_{n=1}^\infty (-1)^n \bar{c}_n e^{in\psi} \bar{\alpha}^{n-1}.
	\end{equation}

This calculation reveals what is so special about the Poisson submanifold.   Recall from Section~\ref{Invariant Manifold of Poisson Kernels} that Poisson kernels arise when we take $\mu$ to be the uniform measure.  Then $c_n = 0$ for all $n \neq 0$ and $\langle z \rangle = \alpha$.  In this exceptional case, the centroid simply evolves according to the Riccati equation (\ref{riccati}) and the dynamics of $\alpha$ and $\psi$ decouple in Eqs.~(\ref{7a}), (\ref{7b}).  (A similar observation about the crucial role of the uniform measure here was made by Pikovsky and Rosenblum~\cite{piko08}.  The centroid evolution equation (\ref{7a}) on the Poisson submanifold was first written down by Ott and Antonsen; see Eq.(6) in Ref.~\cite{otta08}.)  

But for the generic case of states lying off the Poisson submanifold, $\langle z \rangle$ is no longer equal to  $\alpha$ and the reduced dynamics become fully three-dimensional, due to the coupling between $\alpha$ and $\psi$ induced by the relation (\ref{25}) and the dependence of $f$ and $g$ on $\langle z \rangle$ and the higher Fourier coefficients.  In the next section we will explore some of the possibilities for such three-dimensional flows.

	\begin{figure}
	\includegraphics[scale = 1]{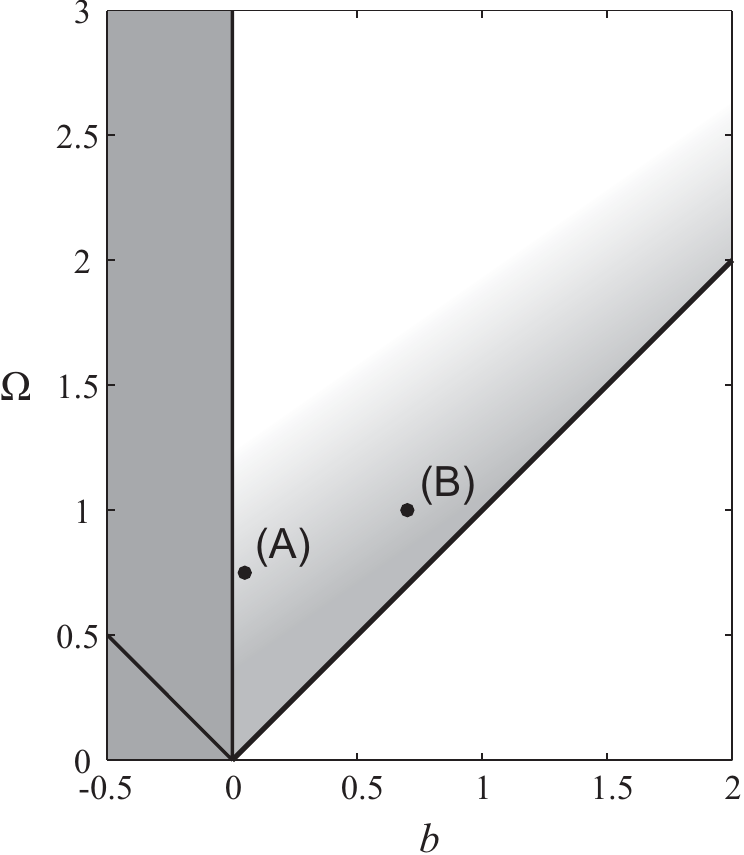}
	\caption{The qualitative trend of chaos observed in the first quadrant of the $b$-$\Omega$ parameter plane is indicated by the shaded gradient.  As the shade darkens near the bifurcation curve $\Omega = b$, chaos fills increasingly larger regions of the submanifolds containing the sinusoidal initial distributions.  Points (A) and (B) are chosen as (1/20, 3/4) and (17/10, 1), respectively.  Representative Poincar\'{e} sections for these points are shown in Fig.~\ref{kamchaosA} and Fig.~\ref{kamchaosB}.  The region $b < 0$ is grayed out to represent that negative values of $b$ are not physical.  \label{pointplot}}
	\end{figure}
	
	\begin{figure*}
	\includegraphics[scale = 1]{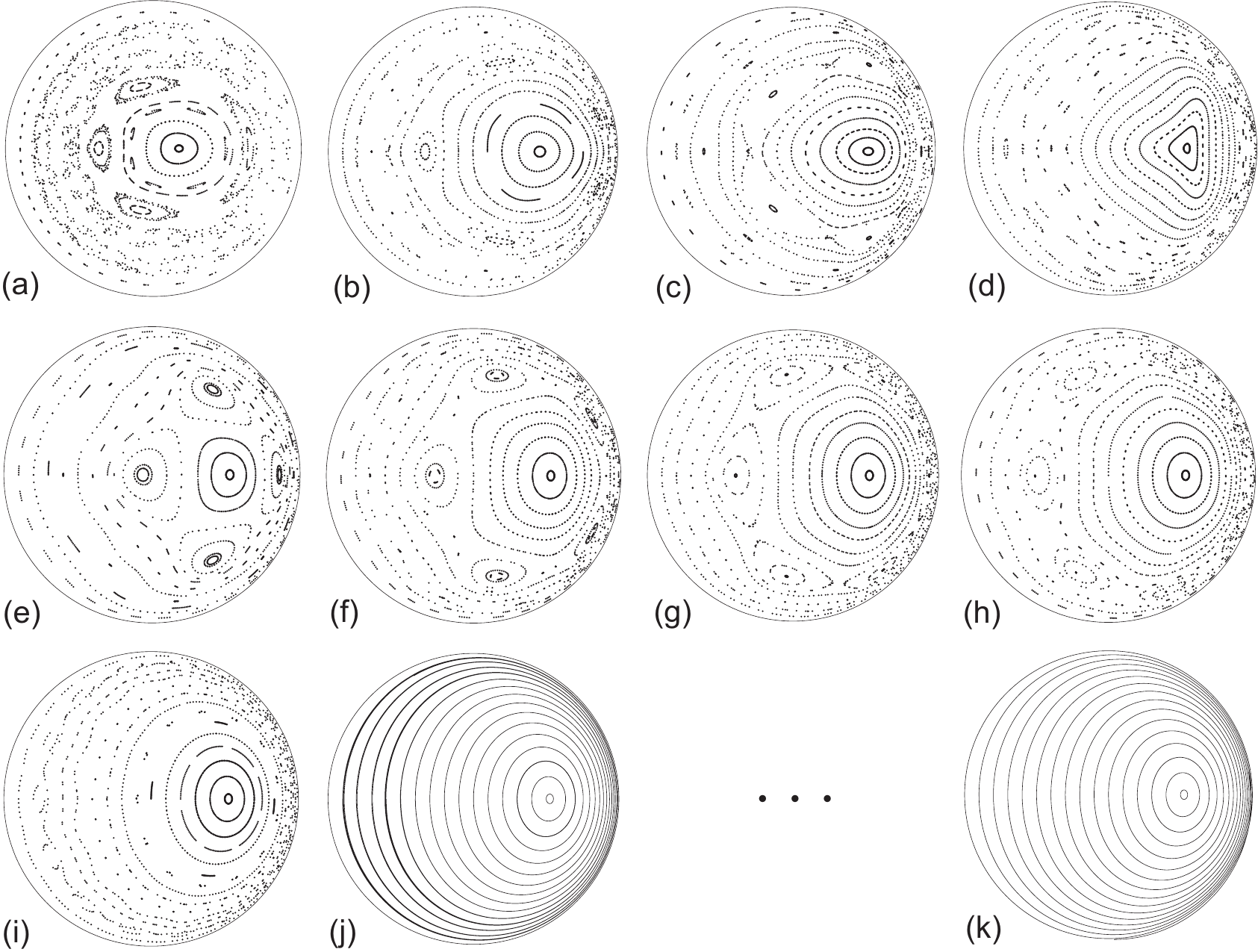}
	\caption{Poincar\'{e} sections of $\alpha$ at $\psi \, (\text{mod} \; 2\pi) = 0$ for a resistively-loaded series array of Josephson junctions with $b = 1/20, \Omega = 3/4$ (pt. (A) in Fig.~\ref{pointplot}).  The initial distributions are sinusoidal with wavenumber $n$, where $n$ is (a) 1, (b) 2, (c) 3, (d) 4, (e) 5, (f) 6, (g) 7, (h) 8, (i) 16, (j) 32, and (k) $\infty$, i.e. on the Poisson submanifold.  In (j) and (k), the complete trajectories are plotted instead of the intersections with the plane $\psi \, (\text{mod} \; 2\pi) = 0$. \label{kamchaosA}}
	\end{figure*}
	
	\begin{figure*}
	\includegraphics[scale = 1]{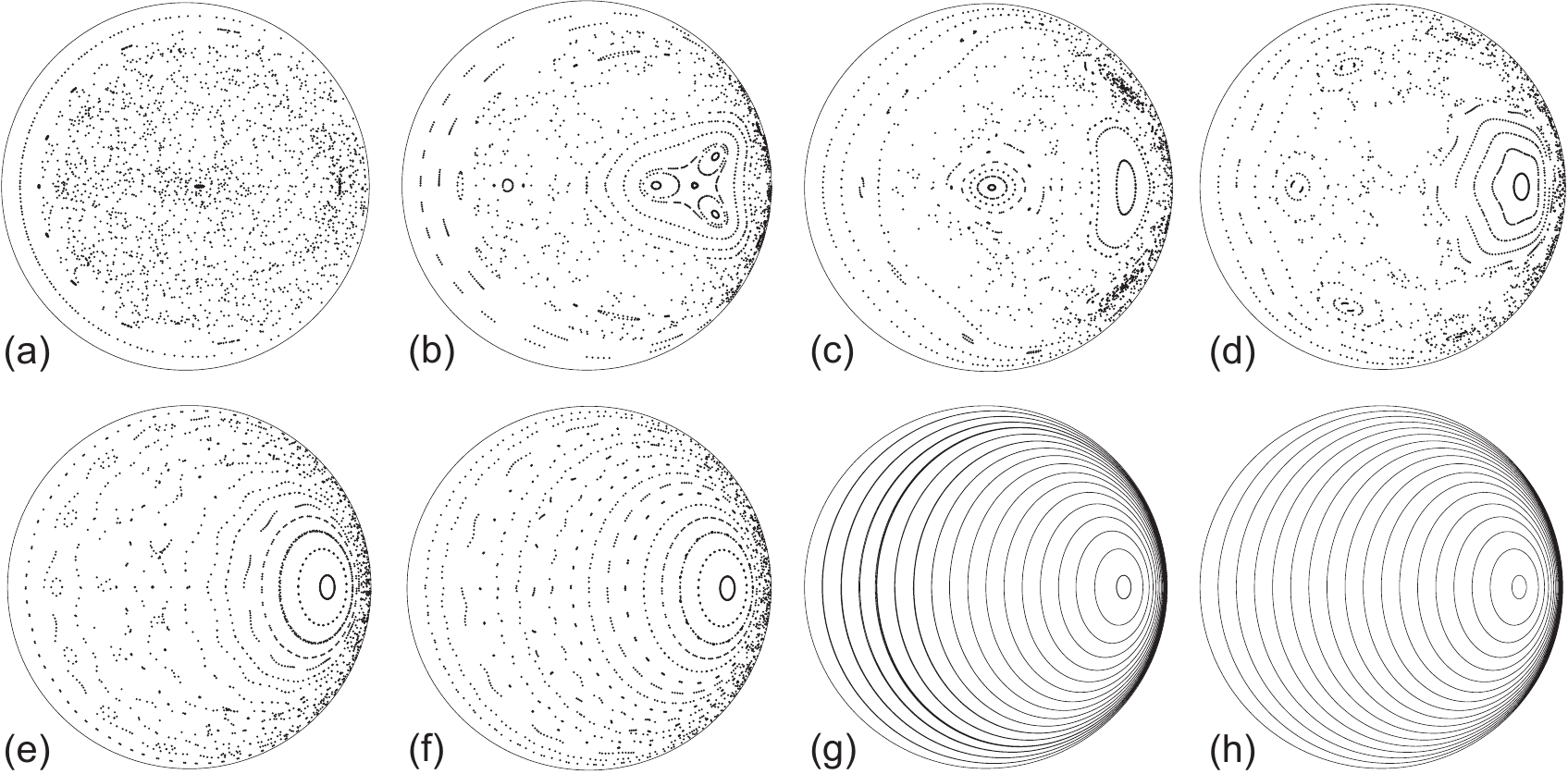}
	\caption{Poincar\'{e} sections of $\alpha$ at $\psi \, (\text{mod} \; 2\pi) = 0$ for a resistively-loaded series array of Josephson junctions with $b = 17/10, \Omega = 1$ (pt. (B) in Fig.~\ref{pointplot}).  The initial distributions are sinusoidal with wavenumber $n$, where $n$ is (a) 1, (b) 2, (c) 4, (d) 8, (e) 16, (f) 32, (g) 64, and (h) $\infty$, i.e. on the Poisson submanifold.  In (g) and (h), the full trajectories are plotted. \label{kamchaosB}}
	\end{figure*}

\section{CHAOS IN JOSEPHSON ARRAYS \label{CHAOS IN JOSEPHSON ARRAYS}}

Although the leaves of the foliation imposed by the M\"{o}bius group action are only three-dimensional, they often contain chaos for commonly studied $f$ and $g$~\cite{golo92, wata94}.  In this section, we showcase this phenomenon by specializing to the case of a resistively-loaded series array of overdamped Josephson junctions.  

In several previous studies of sinusoidally coupled oscillators in the continuum limit, it was found that under certain conditions, the Fourier harmonics of the phase density $\rho(\phi,t)$ evolved as if they were decoupled, at least near certain points in state space~\cite{stro91,golo92,stro93}.  In the spirit of these observations, we can get a sense for how individual harmonics contribute to the chaos by starting the system  (\ref{7}) from sinusoidal phase densities with different wavenumbers $n$.  

To be more precise, we choose an initial density 
 	\begin{equation} \label{initial_rho_density}
	\rho(\phi,0) = \frac{1}{2\pi}(1 + \cos n\phi).
	\end{equation}
At $t=0$, we choose $\alpha = \psi = 0$ so that $M_t$ in Eq.~(\ref{2}) is simply the identity map, and the time-dependent change of variables $e^{i\phi} = M_t (e^{i\theta})$ reduces to $\phi = \theta$, initially. Thus, the corresponding density of $\theta$ is 
 	\begin{equation} \label{26}
	\sigma_n(\theta) = \frac{1}{2\pi}(1 + \cos n\theta).
	\end{equation}
This density is independent of time, just as the angles $\theta_j$ were in the finite-$N$ case. 

Next we flow the density forward by $e^{i\phi} = M_t (e^{i\theta})$, where the M\"{o}bius parameters $\alpha(t), \psi(t)$ satisfy the reduced flow (\ref{7}).  Then, by our earlier results, the resulting density $\rho(\phi,t)$ automatically satisfies the governing equations (\ref{continuity}), (\ref{velocity}).  The three-dimensional plot of $\text{Re} (\alpha(t))$, $\text{Im} (\alpha(t))$ and $\psi(t)$ indicates how such a single-harmonic density evolves in time, revealing for example whether it exhibits chaos, follows a periodic orbit, or approaches a fixed point.

To ease the notation, from now on we write $\alpha$ in Cartesian coordinates as 
 	\begin{equation} \label{Cartesian}
	\alpha = x + iy.  
	\end{equation}
Then the reduced flow (\ref{7}) becomes
	\begin{equation} \label{28}
	\begin{split}
	\dot{x}    &= -uy + \text{Im} (f) (1 - x^2 - y^2) \\
	\dot{y}    &=  ux + \text{Re} (f) (1 - x^2 - y^2) \\
	\dot{\psi} &=  u
	\end{split}
	\end{equation}
where 
 	\begin{equation} \label{u_definition}
	u = 2 x \,\text{Re} (f)  + g - 2 y \, \text{Im} (f).  
	\end{equation}

We immediately see that for every fixed point of this system, $|\alpha| = 1$ and $\psi$ is arbitrary.  If for some change of state variables $\zeta(x,y,\psi)$, $\eta(x,y,\psi)$, and $\xi(x,y,\psi)$, the ODEs $\dot{\zeta}$ and $\dot{\eta}$ constitute a closed two-dimensional system and $\dot{\xi}$ receives all of its $t$-dependence through $\zeta$ and $\eta$, then there could be other fixed points for the physical system, namely where $\dot{\zeta} = \dot{\eta} = 0$ but $\dot{\xi} \neq 0$.  Examples of the second type of fixed point include the splay states found on the Poisson submanifold~\cite{tsan91, stro93}.

As discussed in Section~\ref{Reducible systems}, series arrays of Josephson junctions with a resistive load have dynamics given by Eqs.~(\ref{jj_resistive}), (\ref{continuity}), and (\ref{velocity}), with $f = -(b+1)/2$ and $g = \Omega + \text{Re} \langle z \rangle$, where $b$ and $\Omega$ are dimensionless combinations of certain circuit parameters~\cite{tsan91, marv09} and $ \langle z \rangle$ is the complex order parameter (\ref{z}).  The dynamics of $x$, $y$ and $\psi$ are given by substitution into (\ref{28}):
	\begin{equation} \label{29}
	\begin{split}
	\dot{x}    &= -uy                                \\
	\dot{y}    &=  ux - \frac{b+1}{2}(1 - x^2 - y^2) \\
	\dot{\psi} &=  u	
	\end{split}
	\end{equation}
with $u = \Omega + \text{Re} \langle z \rangle - (b+1)x$.  From (\ref{25}) and (\ref{26}), 
$\text{Re} \langle z \rangle = x + (-1)^n \frac{1}{2}(x^2 + y^2 - 1)(x^2+y^2)^{(n-1)/2} \cos[n\psi-(n-1)\tan^{-1}(y/x)]$.

We can now plot the phase portrait for (\ref{29}) on the cylinder $\{(x,y,\psi) | \, x,y,\psi \in \mathbb{R}, x^2 + y^2 \leq 1\}$.  In the simple case where $\alpha$ decouples from $\psi$, trajectories can be projected down onto the $\alpha$-disc without intersecting themselves or each other.  However, in the more typical case that $\alpha$ and $\psi$ are interdependent, we use Poincar\'{e} sections at $\psi \, (\text{mod} \; 2\pi) = 0$ to sort out the structure.  In these Poincar\'{e} sections, quasiperiodic trajectories (ideally) appear as closed curves or island chains, periodic trajectories appear as fixed points or period-$p$ points of integer period, and chaotic trajectories fill the remaining regions of the unit disc.

First, however, we must choose an appropriate $b$ and $\Omega$.  To do so, we consider their definitions in terms of the original circuit parameters:  $b = R/(NR_J)$ and $\Omega = bI_b/I_c$, where $N$ is the number of junctions, $I_b$ the source current, $R$ the load resistance, $I_c$ the critical current of each Josephson junction, and $R_J$ the intrinsic Josephson junction resistance~\cite{tsan91,marv09}.  Because the resistances must be positive in the physical system, we examine only $b > 0$ in our simulations.  Additionally, $I_c$ represents a positive current magnitude, while $I_b$ reflects both a source current magnitude and direction.  Since the circuit is symmetric with respect to reversal of the source circuit (see Fig. 1 of~\cite{marv09}), the corresponding dynamical system is left unchanged by the reflection $\Omega \rightarrow -\Omega, x \rightarrow -x$.  Hence, we also restrict our study to positive values of $\Omega$.

If $b/\Omega > 1$, (\ref{29}) implies there are fixed points at $x^* = \Omega/b$, $y^* = \pm \sqrt{1 - \Omega^2/b^2}$ for arbitrary $\psi$.  In numerical experiments, the negative-$y^*$ line of fixed points appears to attract distributions, while the positive-$y^*$ line repels them.  Along the bifurcation curve $\Omega = b$, the two rows of fixed points merge at $x = 1$, and we find computational evidence that a splay state (for which $\dot{x} = \dot{y} = 0$) emerges from their union and moves inside the unit disc along the $x$-axis toward the origin as $b$ is decreased or $\Omega$ is increased.  We can see from (\ref{29}) that any such state must lie on the $x$-axis for all parameter values, as it did in previous characterizations of the Poisson submanifold~\cite{marv09}.

For the submanifolds we examined, chaos only appeared in the portion of the first quadrant in the $b$-$\Omega$ plane that did not contain the fixed points, and the chaos became more widespread as $b/\Omega \rightarrow 1$.  This is illustrated schematically in Fig.~\ref{pointplot}; the gradient of increasing darkness represents increasingly pervasive chaos.  In submanifolds where the chaos was not widespread, the dynamics on the Poincar\'{e} sections were reminiscent of a Kolmogorov-Arnold-Moser Hamiltonian system with hierarchies of islands enclosing nested sets of closed orbits.  Nevertheless, we do not have an explicit Hamiltonian for (\ref{jj_resistive}) as we do for its averaged counterpart~\cite{wata93}.

The increase in chaotic behavior is clearly visible in Figs.~\ref{kamchaosA} and Fig.~\ref{kamchaosB}, which show sequences of Poincar\'{e} sections corresponding to the points (A) and (B) in Fig.~\ref{pointplot}.  Point (A) lies at $(b,\Omega) = (1/20,3/4)$, about 1/2 unit from the bifurcation curve $\Omega = b$, while point (B) lies at $(b,\Omega) = (17/10,1)$, about 1/3 unit from $\Omega = b$.  As an example of the pattern of escalating chaos, observe that Figs.~\ref{kamchaosB}(a),(b),(c) have larger, more dramatically overlapping chaotic regions than the corresponding plots (a),(b),(d) of Fig.~\ref{kamchaosA}.

Although not shown, the chaotic trajectories that produced the scattered points in the Poincar\'{e} sections are phase coherent:  they cycle smoothly and unidirectionally around the splay states throughout each period of $\psi$.  When the splay states are moved toward the edge of the unit disc by increasing $b$ or decreasing $\Omega$, these trajectories appear increasingly less prone to return to the same neighborhoods in the Poincar\'{e} sections, resulting in the observed amplification of chaotic behavior.

It is also possible to interpret the association between the parameter values and the intensity of the chaos in terms of the underlying physical parameters.  In terms of these parameters, the limit $b/\Omega \rightarrow 1^-$ translates to $I_c/I_b \rightarrow 1^-$ or $I_b \rightarrow I_c^+$, which predicts that chaos should appear in real series arrays of Josephson junctions if the source current is reduced to near the critical current of the junctions.

Even though the Poincar\'{e} sections in Fig.~\ref{kamchaosA} and Fig.~\ref{kamchaosB} show differing degrees of chaos, both series of plots depict a trend of decreasing chaotic behavior with increasing $n$.  This stems from the dependence of $g$ on the phase centroid $\langle z \rangle$, which in turn arises because the oscillators are coupled only through their effect on the first harmonic of the phase density.  For a coupling of this type, a sinusoidal phase density with a very short period and rapid oscillations (high $n$)  ``looks'' nearly identical (in the Riemann-Lebesgue sense) to a uniform density.  Hence, in the limit of large $n$, we see $\alpha$ decoupling from $\psi$, just as it does on the Poisson submanifold (recall that the Poisson submanifold corresponds to a uniform density in $\theta$, as shown in Section~\ref{Invariant Manifold of Poisson Kernels}).  From this perspective, then, chaos becomes increasingly dominant as we move ``away'' from the Poisson submanifold, down toward small $n$.

Finally, we point out a surprising feature in the Poincar\'{e} sections of (A) that was common in other simulations we performed.  Starting at $n = 5$, we see prominent sets of period-$(n - 1)$ islands which appear for $n$ up to 8 in Fig.~\ref{kamchaosA}.  This ring of islands appears for higher $n$ as well and forms an increasingly larger and thinner band as $n$ is increased.  Inside the dilating band, a set of nested orbits resembling the corresponding neutrally stable cycles of the Poisson submanifold grows, filling the unit disc and approaching coincidence with the trajectories on the Poisson submanifold.  We are currently unclear on why exactly $(n-1)$ islands emerge from the M\"{o}bius group action on (\ref{26}), but pose this as an open question for future study.

Although it is tempting to try to extrapolate our numerical results to the case of non-identical oscillators, Ott and Antonsen~\cite{otta09} have recently demonstrated that such systems contain a two-dimensional submanifold (the generalization of the simpler Poisson submanifold studied here) that carries all the long-term dynamics of the phase centroid $\langle z \rangle$.  Their results hold for the common case in which $g$ is a time-independent angular frequency with some distribution of values among the oscillators, and $f$ is a function of time, independent of oscillator variability.  Our numerical experiments, together with this new result, indicate that the widespread neutral stability in systems of identical, sinusoidally-coupled phase oscillators is a consequence of their special symmetries and underlying group-theoretic structure.

\section{APPENDIX \label{APPENDIX}}

We show that the $N!/(N-4)!$ cross ratios of the oscillator phases are functionally dependent on the $N - 3$ cross ratios $\{\lambda_{1234}, \lambda_{2345}, \dotsc, \lambda_{(N-3)(N-2)(N-1)N}\}$.  To do so, we use the fact that the $4!$ cross ratios corresponding to the $4!$ permutations of $z_i, z_j, z_k, z_l$ can be written as elementary functions of $\lambda_{ijkl}$:
	\begin{equation} \label{14.2}
	\begin{split}	
	\lambda_{ijkl} &= \lambda_{jilk} = \lambda_{klij} = \lambda_{lkji} \\
	\lambda_{ijlk} &= 1/\lambda_{ijkl}                                 \\
	\lambda_{iklj} &= 1/(1 - \lambda_{ijkl})                           \\
	\lambda_{ikjl} &= 1 - \lambda_{ijkl}                               \\
	\lambda_{ilkj} &= \lambda_{ijkl}/(1 - \lambda_{ijkl})              \\
	\lambda_{iljk} &= (\lambda_{ijkl} - 1)/\lambda_{ijkl}              \\
	\end{split}
	\end{equation}
Additionally, we can obtain new cross ratios from existing ones by multiplication:
	\begin{equation} \label{15}
	\lambda_{ijkl}\lambda_{jmkl} = \lambda_{imkl}  
	\end{equation}

Using these facts, we need to show that we can write $\lambda_{PQRS}$ for any distinct $P, Q, R, S \in \{1, 2, \dotsc, N\}$ in terms of elements from $\{\lambda_{1234}, \lambda_{2345}, \dotsc, \lambda_{(N-3)(N-2)(N-1)N}\}$.  First, note that we can rewrite (\ref{15}) as a function $F_j$ which takes two cross ratios $\lambda_{ijkl}$ and $\lambda_{jklm}$ (with indices in order), permutes the indices as necessary to eliminate $z_j$, executes the multiplication and returns the product with its indices in order:
	\begin{equation} \label{16}
	F_j(\lambda_{ijkl},\lambda_{jklm}) = \lambda_{iklm}
	\end{equation}
Observe, however, that $F_j$ is just short-hand for a composition of elementary functions from (\ref{14.2}):
	\begin{equation} \label{17}
	F_j(\lambda_{ijkl},\lambda_{jklm}) = \frac{1}{1 - \lambda_{ijkl}(\lambda_{jklm} - 1)/\lambda_{jklm}}
	\end{equation}
We can also define the analogous functions $G_k$ and $H_l$:
	\begin{equation} \label{30}
	\begin{split}
	G_k(\lambda_{ijkl},\lambda_{jklm}) &= \lambda_{ijlm} \\
	H_l(\lambda_{ijkl},\lambda_{jklm}) &= \lambda_{ijkm} \\
	\end{split}
	\end{equation}
These functions have their own compositions like that of $F_j$ in (\ref{17}).

Let $\lambda_{pqrs}$ correspond to the permutation of $\lambda_{PQRS}$ in which the indices are in order.  We can write $\lambda_{PQRS}$ in terms of $\lambda_{pqrs}$ using one of the functions in (\ref{14.2}).  Thus, the problem reduces to showing that we can obtain $\lambda_{pqrs}$ from the elements of $\{\lambda_{1234}, \lambda_{2345}, \dotsc, \lambda_{(N-3)(N-2)(N-1)N}\}$ by elimination of the indices between $p$, $q$, $r$, $s$ using the operations $F_j$, $G_k$, $H_l$.  

If there are one or more indices between $i$ and $j$, we say there is a \textit{gap} between $i$ and $j$.  Now observe that we can obtain the first gap between $p$ and $q$ using only $\lambda_{ijkl}$ with no gaps; we grow this gap iteratively one index at a time by the operation:  $F_k(\lambda_{pk(k+1)(k+2)},\lambda_{k(k+1)(k+2)(k+3)}) = \lambda_{p(k+1)(k+2)(k+3)}$.  We can then grow the second gap between $q$ and $r$ to its full size using only $\lambda_{ijkl}$ that have no gaps between $j$ and $k$ or $k$ and $l$ (each of which could be made from $\lambda_{ijkl}$ with no gaps) using the operation:  $G_k(\lambda_{pqk(k+1)},\lambda_{qk(k+1)(k+2)}) = \lambda_{pq(k+1)(k+2)}$.  Finally, we can create the third gap between $r$ and $s$ using only $\lambda_{ijkl}$ with no gaps between $k$ and $l$ (which could be made from $\lambda_{ijkl}$ with fewer gaps) using the operation:  $H_k(\lambda_{pqrk},\lambda_{qrk(k+1)}) = \lambda_{pqr(k+1)}$.

Since each $\lambda_{ijkl}$ (with $i < j < k < l$) can be built up from $\lambda_{ijkl}$ with fewer gaps, the proof is complete:  all $N!/(N-4)!$ cross ratios are dependent on the elements of $\{\lambda_{1234}, \lambda_{2345}, \dotsc, \lambda_{(N-3)(N-2)(N-1)N}\}$.

\vspace{10 pt}

\textbf{Acknowledgments:}
Research supported in part by National Science Foundation grant NSF CISE-0835706.

\end{document}